\documentclass[documentclass[runningheads]{llncs}
\usepackage{fancyhdr,graphicx,amsmath,amssymb}
\usepackage{inputenc}
\usepackage[ruled,vlined]{algorithm2e}
\usepackage{algorithmic}
\usepackage{float}
\usepackage{upgreek}
\usepackage{subfig}
\usepackage{array}
\usepackage{multirow}
\usepackage{amssymb}
\usepackage{amsmath, mathtools}
\usepackage{graphicx}
\usepackage{calc}
\usepackage[section]{placeins}
\usepackage{listings}
\usepackage{xcolor}
\usepackage{url}
\usepackage{enumitem}
\usepackage{textcomp}
\usepackage{scalerel}
\usepackage{booktabs}
\urldef{\mailsb}\path|jcarrasquel@hse.ru|
\urldef{\mailsa}\path|k_mecheraoui@esi.dz|
\urldef{\mailsc}\path|ilomazova@hse.ru|
\newcommand{\keywords}[1]{\par\addvspace\baselineskip
\noindent\keywordname\enspace\ignorespaces#1}
\usepackage{eurosym}
\def\code#1{\textup{\texttt{#1}}}
\AtBeginDocument{}

\usepackage{nomencl}
\makenomenclature
\nomenclature{$\exists"!$}{exists exactly one}

\begin{document}

\mainmatter
\title{Searching for Deviations in Trading Systems: Combining Control-Flow and Data Perspectives\thanks{This work is supported by the Basic Research Program at the National Research University Higher School of Economics.}}
\titlerunning{Object-Centric Conformance Checking and its Application on Trading Systems}
\author{Julio C. Carrasquel and Irina A. Lomazova}
\institute{HSE University\\
Myasnitskaya ul. 20, 101000 Moscow, Russia\\ \mailsb , \mailsc}

\maketitle

\begin{abstract}
Trading systems are software platforms that support the exchange of securities (e.g., company shares) between participants. In this paper, we present a method to search for deviations in trading systems by checking conformance between colored Petri nets and event logs. Colored Petri nets (CPNs) are an extension of Petri nets, a formalism for modeling of distributed systems. CPNs allow us to describe an expected causal ordering between system activities and how data attributes of domain-related objects (e.g., orders to trade) must be transformed. Event logs consist of traces corresponding to runs of a real system. By comparing CPNs and event logs, different types of deviations can be detected. Using this method, we report the validation of a real-life trading system.
\end{abstract}

\begin{section}{Introduction}
Trading systems are software platforms that support the exchange of securities (e.g., company shares) between participants \cite{Harris2003}. In these systems, orders are submitted by users to indicate what securities they aim to buy or sell, how many stocks and their price. Investors buy securities with promising returns, whereas companies sell their shares to gain capital. These are some of the reasons why trading systems are a vital element in global finances, requiring software processes in these systems to guarantee their correctness. Among these processes, a crucial one is the management of orders in order books. Order books are two-sided priority lists where buy orders and sell orders that aim to trade the same security are matched for trading. A trading system must handle and match orders in these books according to its specification. Nonetheless, trading systems may be prone to deviate from their specification due to software errors or malicious users. This is why the validation of processes in trading systems, such as the management of orders, is a task of utmost importance. In this light, domain experts constantly seek for novel ways to detect system \emph{deviations}, that is, to localize precise differences between a real system and its specification \cite{Itkin2019}.

\noindent To detect deviations in trading systems, we consider \emph{conformance checking} \cite{vanDerAalst2016,Carmona2018}. Conformance checking is a process mining technique to search for differences between models describing \emph{expected behavior} of processes and event logs that record \emph{real behavior} of such processes \cite{vanDerAalst2016}. Event logs consist of traces related to runs of processes; a trace is a sequence of events, where each event indicates an activity executed. To model expected behavior, we use \emph{Petri nets} --- a well-known formalism for modeling distributed systems \cite{Murata1989}. Petri nets allow to describe the \emph{control-flow} perspective of processes, that is, activities and their causal ordering, e.g., ``a trade between two orders is preceded by the submission of both orders".

For trading systems, models should describe not only control-flow, but also how \emph{data attributes} of objects such as orders change upon the execution of activities (e.g., ``stocks of a sell order decrease by the number of stocks sold in a trade"). We resort to colored Petri nets (CPNs) to combine both control-flow and data perspectives \cite{Jensen2009}. CPNs are an extension of Petri nets where tokens progressing through the net carry data from some domains (referred to as ``colors"). CPNs allow us to describe how trading systems handle objects (represented by tokens) and how their data attributes are transformed. This is an advantage over data-aware Petri net models used in other conformance methods, which do not directly relate data to tokens \cite{Mannhardt2015}. In \cite{Carrasquel2019-2,Carrasquel2019,PNSE} we presented how CPNs, as well as other Petri net extensions, allow to model different processes in trading systems. 

We then developed conformance methods to replay traces of trading systems on CPNs. Replay comprises the execution of a model based on the information in events of a trace \cite{Rozinat2008-2}. Deviations are found when a model cannot be executed as an event indicates. In \cite{Carrasquel2021} we consider deviations related only to control-flow, proposing a strategy to force the model execution when such deviations are found. In \cite{AIST2020} we use replay to check if data attributes of objects are transformed by a real system in the same way that its model does. Yet, the method in \cite{AIST2020} does not use any strategy to force the model execution if deviations are found, thereby halting the replay upon the first occurrence of a deviation in a trace.

This paper presents a comprehensive conformance method that integrates and extends the approaches presented in \cite{Carrasquel2021,AIST2020} in order to detect multiple kinds of deviations, including those related to control-flow and data attributes of objects. Notably, strategies are provided to force the execution of a CPN upon the occurrence of each kind of deviation. In particular, the following kinds of deviations can be detected when replaying a system's trace on a CPN that models the system specification: $(i)$ \emph{control-flow deviation:} the real system invoked an activity involving certain objects, but skipping some activities that should have been executed before to handle such objects; $(ii)$ \emph{priority rule violation:} an object was served before other objects with higher priority; $(iii)$ \emph{resource corruption:} object attributes were not transformed as the model specifies; $(iv)$ \emph{non-proper termination:} an object was not fully processed by the real system. The method returns a file with precise information about all deviations detected. We developed a prototypical implementation of the method, that we use to validate the management of orders in a real trading system. An experiment with artificial data is also reported.

The remainder of this paper is structured as follows. Sections 2 and 3 introduce the CPN models and event logs used in our method. Section 4 presents the conformance method. Section 5 reports the prototype and experiments conducted. Finally, Section 6 presents the conclusions.

\end{section}

\begin{section}{Colored Petri Nets}

Petri nets \cite{Murata1989} are bipartite graphs consisting of two kinds of nodes: places and transitions. Places (drawn as circles) denote resource buffers, conditions, or virtual/physical locations. Transitions (drawn as boxes) account for system activities. Places store tokens, denoting control threads, resources, etc. Transitions consume tokens from input places and produce them in output places. We consider colored Petri nets (CPNs), where tokens carry data belonging to some data domains (``colors") \cite{Jensen2009}. Fig. \ref{fig:cpn-model} depicts a CPN modeling a trading system handling buy/sell orders in one order book. Places $p_1$ and $p_2$ are sources for incoming buy and sell orders; $p_3$ and $p_4$ are buffers for submitted orders; $p_5$ and $p_6$ model the buy/sell side of the order book, whereas $p_7$ and $p_8$ are sinks for orders that traded or were canceled. Transitions $t_1$ and $t_2$ model submission of orders by users; $t_3$ and $t_4$ model insertion of orders in the order book. Transition $t_5$ (activity \code{trade1}) models a trade where two involved orders are filled (all their stocks were bought/sold); $t_6$, $t_7$ (activities \code{trade2} and \code{trade3}) model the cases where only one order is filled, whereas the second one is partially filled (returning to the order book). Transitions $t_8$ and $t_9$ model cancellation of orders.

\begin{figure}[H]
\vspace*{-\baselineskip}
\centering
\includegraphics[scale=0.701]{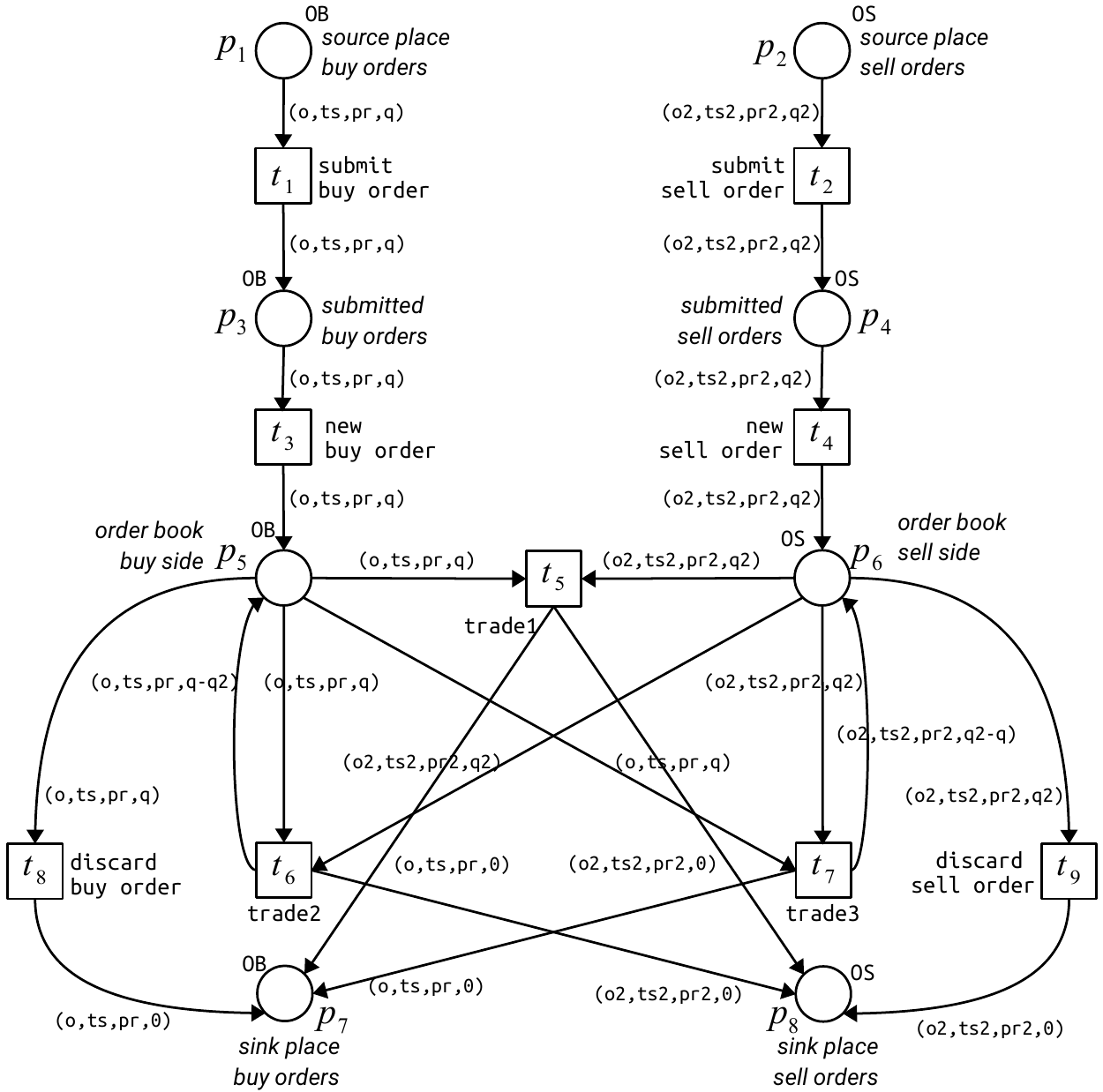}
\caption{CPN model of a trading system operating one order book.}
\label{fig:cpn-model}
\vspace*{-\baselineskip}
\end{figure}

\noindent Let $\mathfrak{D}$ be a finite set of \emph{data domains}. A Cartesian product $D_1 \times ... \times D_n$, $n \geq 1$, between a combination of data domains $D_1, ... , D_n$ from $\mathfrak{D}$ is called a \textit{color}. $\Sigma$ is the set of all possible colors defined over $\mathfrak{D}$. A token is a tuple $(d_1, ..., d_n) \in \code{C}$ s.t. $\code{C}$ is a color in $\Sigma$, and we call the first component $d_1$ as the token's \emph{identifier}. In Fig. 1, we have colors $\code{OB} = O_{\code{B}} \times \mathbb{N} \times \mathbb{R}^{+} \times \mathbb{N}$ and $\code{OS} = O_{\code{S}} \times \mathbb{N} \times \mathbb{R}^{+} \times \mathbb{N}$,  where $O_{\code{B}}$ and $O_{\code{S}}$ are sets of identifiers for buy orders and sell orders, $\mathbb{N}$ is the set of natural numbers (including zero) and $\mathbb{R}^{+}$ is the set of positive real numbers; these colors denote orders with identifiers, arrival time, price and stock quantity, e.g., a token $(\code{b1},\code{1},\code{22.0},\code{5})$ denotes a buy order with identifier $\code{b1}$, submitted in time $\code{1}$, to buy $\code{5}$ stocks at price $\code{22.0}$ per stock. Thus, colors model classes of objects, whereas tokens are object instances. We fix a function $\code{color}$ to indicate the color of tokens that each place stores (e.g., $\code{color}(p_1) = \code{OB}$). Arcs are labeled with expressions to specify how tokens are processed. We fix a language of expressions $\mathcal{L}$, where each expression is of the form $(e_1,...,e_n)$ s.t., for each $i \in \{1,...,n\}$, $e_i$ is either a constant from a domain in $\mathfrak{D}$, a variable typed over an element in $\mathfrak{D}$, or a function whose domain and range are domains in $\mathfrak{D}$. For a variable $\code{v}$ we denote its type by \code{type}$($\code{v}$)$, s.t. \code{type}$($\code{v}$) \in \mathfrak{D}$. With slight abuse of notation, for an expression $(e_1,...,e_n)$ we have that $\code{color}((e_1,...,e_n)) = D_1 \times ... \times D_n$ where, for each $i \in \{1,...,n\}$, $D_i \in \mathfrak{D}$ and $D_i = \code{type}(e_i)$ if $e_i$ is a variable, $e_i \in D_i$ if $e_i$ is a constant, or $D_i$ is the range of $e_i$ if $e_i$ is a function.

\begin{definition}[Colored Petri net]
Let $\mathfrak{D}$ be a finite set of data domains, let $\Sigma$ be a set of colors defined over $\mathfrak{D}$, let $\mathcal{L}$ be a language of expressions, and let $\mathcal{A}$ be a set of activity labels. A colored Petri net is a 6-tuple $CP = (P, T, F, \code{color}, \mathcal{E}, \Lambda)$, where:
\begin{itemize}[leftmargin=*]
\item $P$ is a finite set of places, $T$ is a finite set of transitions, s.t. $P \cap T = \emptyset$, and $F \subseteq (P \times T) \cup (T \times P)$ is a finite set of directed arcs;
\item $\code{color}: P \rightarrow \Sigma$ is a place-coloring function, mapping each place to a color;
\item $\mathcal{E} : F \rightarrow \mathcal{L}$  is an arc-labeling function, mapping each arc $r$ to an expression in $\mathcal{L}$, such that $\code{color}(\mathcal{E}(r)) = \code{color}(p)$ where $p$ is adjacent to $r$;
\item $\Lambda : T \rightarrow \mathcal{A}$ is an activity-labeling function, mapping each transition to an element in $\mathcal{A}$, $\forall t, t^{\prime} \in T$ : $t \neq t^{\prime} \iff \Lambda(t) \neq \Lambda(t^{\prime})$.
\end{itemize}
\end{definition}

\noindent In the following, for a transition $t \in T$ in a CPN, $\prescript{\bullet}{}{t} = \{ p \in P \; | \; (p,t) \in F \} $ denotes the set of \emph{input places} of $t$, and $t^{\bullet}  = \{ p \in P \; | \; (t,p) \in F \} $ denotes the set of \emph{output places} of $t$. Let $CP = (P,T,F,\code{color},\mathcal{E},\Lambda)$ be a CPN. A \emph{marking} $M$ is a function, mapping every place $p \in P$ to a (possibly empty) multiset of tokens $M(p)$ defined over $\code{color}(p)$. We denote by $M_0$ the \emph{initial marking} of a CPN. A \emph{binding} $b$ of a transition $t \in T$ is a function, that assigns a value $b(\code{v})$ to each variable $\code{v}$ occurring in arc expressions adjacent to $t$, where $b(\code{v}) \in \code{type}(\code{v})$. Transition $t$ is \emph{enabled} in marking $M$ w.r.t. a binding $b$ iff $\forall p \in \prescript{\bullet}{}{t} : b(\mathcal{E}(p,t)) \in M(p)$, that is, each input place of $t$ has at least one token to be consumed. The $\emph{firing}$ of an enabled transition $t$ in a marking $M$ w.r.t. to a binding $b$ yields a new marking $M^{\prime}$ such that $\forall p \in P : M^{\prime}(p) =  M(p) \setminus \{ b(\mathcal{E}(p,t)) \} \cup \{ b(\mathcal{E}(t,p)) \}$.

\smallskip

\noindent Finally, we define restrictions for CPNs in order to model processes in trading systems that handle different kinds of objects such as buy/sell orders \cite{Carrasquel2021,AIST2020}. We call \emph{conservative-workflow} CPNs the models that comply with such restrictions.

\begin{definition}[\textbf{Conservative-Workflow Colored Petri Net}]
Let $\Sigma$ be a finite set of colors, let $CP = (P, T, F, \normalfont{\code{color}}, \mathcal{E}, \Lambda)$ be a CPN defined over $\Sigma$, and let $M_{0}$ be the initial marking of $CP$. We say that $CP$ is a conservative-workflow CPN iff:

\begin{enumerate}[leftmargin=*]
    \item CP is a conservative colored Petri net, such that for every transition $t \in T$:
    \begin{itemize}[leftmargin=*]
        \smallskip
        \item $\forall \; p \in \prescript{\bullet}{}{t} \;\; \exists! \; p^{\prime} \in t^{\bullet} : \mathcal{E}(p,t) = (\code{v}_1,...,\code{v}_n) \land \mathcal{E}(t,p^{\prime}) = (\code{w}_1,...,\code{w}_n) \land \; \code{v}_1 = \code{w}_1$.
        \item $\forall \; p \in t^{\bullet} \;\; \exists! \; p^{\prime} \in \prescript{\bullet}{}{t} : \mathcal{E}(p^{\prime},t) = (\code{v}_1,...,\code{v}_n) \land \mathcal{E}(t,p) = (\code{w}_1,...,\code{w}_n) \land \; \code{v}_1 = \code{w}_1$.
    \end{itemize}
    \smallskip
    {\normalfont The restriction above states that for each input arc $(p,t)$ of a transition $t$ with expression $(\code{v}_1,...,\code{v}_n)$, there is exactly one output arc $(t, p^{\prime})$ of $t$ with expression $(\code{w}_1,...,\code{w}_n)$ s.t. $\code{w}_1 = \code{v}_1$; also, for each output arc of $t$ with expression $(\code{w}_1,...,\code{w}_n)$, there is exactly one input arc of $t$ with expression $(\code{v}_1,...,\code{v}_n)$ s.t. $\code{w}_1 = \code{v}_1$; when firing a transition $t$, this restriction guarantees the ``transfer" of a token $(d_1,...,d_n$) from an input place $p$ of $t$ to an output place $p^{\prime}$ with the token's first component $d_1$ unchanged (its ``identifier"); components $d_2,...,d_n$ of the token may be modified by the expression of the output arc $(t, p^{\prime})$ abstractly meaning the transformation of the object represented by the token; thus, tokens with their identifiers cannot ``disappear" or ``duplicate".}
    \vspace{1pt}
    \item There are no two tokens in the initial marking $M_0$ of CP with the same identifier, that is, all tokens have distinct identifiers. Note that if CP is conservative (as defined above), it follows that all tokens have distinct identifiers in every possible marking of $CP$ reachable from the initial marking $M_0$.
    \vspace{1pt}
    \item For every color $\normalfont{\code{C}} \in \Sigma$, there exists one distinguished pair of places in $P$, a source $i$ and a sink $o$, where $\normalfont{\code{color}}(i) = \normalfont{\code{color}}(o) = \normalfont{\code{C}}$, and there exists a path from $i$ to $o$ s.t. for each place $p$ in the path $\normalfont{\code{color}}(p) = \normalfont{\code{C}}$. We respectively denote the sets of source and sink places in $CP$ by $P_0$ and $P_F$.
    \vspace{1pt} 
    \item $\forall t \in T$ : $\; (\forall p, p^{\prime} \in \prescript{\bullet}{}{t} \; p \neq p^{\prime} \iff \normalfont{\code{color}}(p) \neq \normalfont{\code{color}}(p^{\prime})) \; \land \; (\forall p, p^{\prime} \in t^{\bullet} \; p \neq p^{\prime} \iff \normalfont{\code{color}}(p) \neq \normalfont{\code{color}}(p^{\prime}))$, i.e., for every transition $t$, input places of $t$ have distinct colors. The same rule holds for output places of $t$.
\end{enumerate}
\end{definition}

\end{section}

\begin{section}{Event Logs}

\begin{definition}[\textbf{Event, Trace, Event Log}]
\label{def:eventlogs}
Let $\mathfrak{D}$ be a finite set of data domains, let $\Sigma$ be a set of colors defined over $\mathfrak{D}$, and let $\mathcal{A}$ be a finite set of activities. An event is a pair $e = (a, R(e))$ s.t. $a \in A$ and $R(e)$ is a set where $\forall r \in R(e)$, $r \in \code{C}$ and $\code{C} \in \Sigma$. Each element $r$ in $R(e)$ represents an object involved in the execution of activity $a$ in event $e$. A trace $\sigma = \langle e_1,...,e_m \rangle$, $m \geq 1$, is a finite sequence of events. An event log $L$ is a multiset of traces.

\vspace*{-\baselineskip}

\end{definition}

\begin{table}[H]

\vspace*{-\baselineskip}

\caption{A trace $\sigma$ of an event log, corresponding to a run in a trading system.}
\label{tab:table1}
\centering
\begin{tabular}{|c|l|l|}
\hline
\textbf{event} ($e$) & \multicolumn{1}{|c|}{\textbf{activity} ($a$)}& \multicolumn{1}{|c|}{\textbf{objects} ($R(e)$)}\\
\hline
$e_1$ & \code{submit buy order} $\;\;$ & (\code{b1}, \code{1}, \code{22.0}, \code{5})\\

$e_2$ & \code{new buy order} & (\code{b1}, \code{1}, \code{22.0}, \code{5})\\

$e_3$ & \code{submit sell order} & (\code{s1}, \code{2}, \code{21.0}, \code{2})\\

$e_4$ & \code{new sell order} & (\code{s1}, \code{2}, \code{21.0}, \code{2})\\

$e_5$ & \code{new sell order} & (\code{s2}, \code{3}, \code{19.0}, \code{1})\\

$e_6$ & \code{trade2} & (\code{b1}, \code{1}, \code{22.0}, \code{4}), (\code{s1}, \code{2}, \code{21.0}, \code{0})\\
\hline
\end{tabular}

\vspace*{-\baselineskip}

\end{table}

\noindent We denote as $\code{color}(r)$ the color of element $r \in R(e)$ in event $e$. For each object $r = (r^{(1)},...,r^{(n)})$ in an event $e = (a, R(e))$, its components $r^{(1)}, ..., r^{(n)}$ represent the state of $r$ after the execution of $a$. We assume that the first component of $r$, $r^{(1)}$, is the \emph{object identifier} which cannot be modified; $\code{id}(r) = r^{(1)}$ denotes the identifier of $r$. We consider that objects in a trace can be distinguished. $R(\sigma)$ denotes the set of distinct object identifiers in a trace $\sigma$, e.g., for Table 1, $R(\sigma) = \{ \code{b1}, \code{s1}, \code{s2} \}$. Let $r = (r^{(1)},...,r^{(n)})$ be an object. For $j \in \{1,...,n\}$, we consider that each attribute $r^{(j)}$ can be accessed using a name. Objects of the same color share the same set of attribute names, e.g., for color $\code{OB}$ described in Section 2, we consider names $\{\code{id},\code{tsub},\code{price},\code{qty}\}$; we fix a \emph{member access function} $\code{\#}$, that given an object $r = (r^{(1)},...,r^{(n)})$ and the name of the $j$th-attribute, it returns $r^{(j)}$, i.e., $\code{\#}(r,\code{name}_j) = r^{(j)}$.\\For simplicity, we use $\code{name}_j(r)$ instead of $\code{\#}(r,\code{name}_j)$, e.g., for $ r = (\code{b1}, \code{1}, \code{22.0}, \code{5})$, $\code{tsub}(r) = 1$, $\code{price}(r) = \code{22.0}$, and $\code{qty}(r) = \code{5}$.

\smallskip

\noindent Finally, a criterion of \emph{syntactical correctness} must hold for CPNs and event logs that serve as input to the method we propose. Let $L$ be an event log, and let $CP = (P, T, F, \code{color}, \mathcal{E}, \Lambda)$ be a conservative-workflow CPN. We say that $L$ is \emph{syntactically correct} w.r.t. to $CP$ iff, for every trace $\sigma \in L$, each event $e$ in $\sigma$ is syntactically correct. An event $e = (a, R(e))$ is syntactically correct w.r.t. to $CP$ iff $\exists t \in T : \Lambda(t) = a \; \land \; \forall p \in \prescript{\bullet}{}{t} \;\; \exists! r \in R(e) : \code{color}(r) = \code{color}(p) \; \land \;  \forall r \in R(e) \;\; \exists! p \in \prescript{\bullet}{}{t} : \code{color}(r) = \code{color}(p)$; that is, for every event $(a, R(e))$, there exists a transition $t$ with activity label $a$, and each input place of $t$ is mapped to exactly one event's object, and similarly each event's object is mapped to exactly one input place of $t$.

\end{section}

\begin{section}{Conformance Method}

We present a replay-based method to check conformance between a CPN and a trace of an event log. For each event in a trace, the method seeks to execute a model transition labeled with the event's activity, and consumes tokens that correspond to objects involved in the event. As mentioned in Section 1, four kinds of deviations can be detected in events: control-flow deviations, priority rule violations, resource corruptions, and non-proper termination of objects.

\smallskip

\noindent Algorithm 1 describes the replay method between a trace $\sigma$ and a conservative workflow CPN whose initial marking is empty. In addition to deviations, the method returns two counters: the number of \emph{token jumps} $\code{j}$, i.e., the number of tokens that are moved to input places of transitions to force their firing, and the number of consumed/produced tokens $\code{k}$. At the start, each source place of the CPN is populated with the trace's distinct objects $R(\sigma)$ according to their color. For each object to insert as a token in a source place, we set its values according to its first occurrence in $\sigma$. As an example, let us consider the replay of trace $\sigma$ in Table 1 on the CPN of Fig. \ref{fig:cpn-model}: place $p_1$ is populated with buy orders $(\code{b1},\code{1},\code{22.0},\code{5})$, and $p_2$ with sell orders $(\code{s1},\code{2},\code{21.0},\code{2})$ and $(\code{s2},\code{3},\code{19.0},\code{1})$. Then, for each event $e = (a, R(e))$ in $\sigma$, a transition is selected to fire s.t. $\Lambda(t) = a$.

\noindent To fire $t$, we check for every object $r \in R(e)$ whether its corresponding token in the model $(d_1,...,d_n)$, $\code{id}(r) = d_1$, is located in input place $p$ of $t$ s.t. $\code{color}(p) = \code{color}(r)$. If the latter is not true for an object $r$, we look for its corresponding token in other places, which is moved to the input place $p$ of $t$ for tokens of $\code{color}(r)$. In such a case, a \emph{control-flow deviation} is registered and the number of token jumps increases (e.g., Lines 5-10).

\smallskip

\begin{algorithm}[H]
{\footnotesize
	\KwIn{\noindent $CP = (P,T,F,\code{color},\mathcal{E},\Lambda)$ --- conservative-workflow CPN;\\
	\quad \quad \quad $\;P_{0},P_{F} \subseteq P$ --- non-empty sets of source and sink places;\\
	\quad \quad \quad $\; \sigma$ --- an event log trace;}
	\KwOut{counter of token jumps ($\code{j}$) and consumed/produced tokens ($\code{k}$);}	
	\nl $\code{j} \gets \code{0}$; $\code{k} \gets \code{0}$\;
	\nl $\code{populateSourcePlaces}(P_{0},R(\sigma))$\;
	\nl \textbf{foreach} $e = (a, R(e))$ $\code{in}$ $\sigma$ \textbf{do}\\
	\nl \quad \quad $t \gets \code{selectTransition}(a)$\tcp*{$\exists! t \in T \; \Lambda(t) = a$}
    \nl \quad \quad \textbf{foreach} $r$ $\code{in}$ $R(e)$ \textbf{do}\\	
	\nl \quad \quad \quad \textbf{if} $\neg \exists (d_1,...,d_n) \in M(p) : p \in \prescript{\bullet}{}{t} \land \code{color}(p) = \code{color}(r) \land \code{id}(r) = d_1$ \textbf{then}\\
	\nl \quad \quad \quad \quad $\code{registerDeviation}(\code{CONTROL}\_\code{FLOW})$\;
	\nl \quad \quad \quad \quad $\code{jump}(\code{id}(r), p)$\;
	\nl \quad \quad \quad \quad $\code{j} \gets \code{j} + 1$\;
	\nl \quad \quad \quad \textbf{endif}\\
	\nl \quad \quad \quad \textbf{if} $\code{priorityRuleViolation}((d_1,...,d_n), M(p))$ \textbf{then}\\
	\nl \quad \quad \quad \quad $\code{registerDeviation}(\code{RULE}\_\code{VIOLATION})$\;
	\nl \quad \quad \quad \textbf{endif}\\
	\nl \quad \quad \textbf{endfor}\\
	\nl \quad \quad $\code{fire}(t,R(e))$\;
	\nl \quad \quad $\code{k} \gets \code{k} + | R(e) |$\;
	\nl \quad \quad \textbf{foreach} $r$ $\code{in}$ $R(e)$ \textbf{do}\\
	\nl \quad \quad \quad \textbf{let} $d = (d_1,...,d_n) : d_1 = \code{id}(r) \land d \in M(p) \land \code{color}(p) = \code{color}(r)  \land p \in t^{\bullet}$\\
	\nl \quad \quad \quad \textbf{if} $d \neq r$ \textbf{then}\\
	\nl \quad \quad \quad \quad $\code{registerDeviation}(\code{RESOURCE}\_\code{CORRUPTED})$\;
	\nl \quad \quad \quad \quad $d \gets r$\;
	\nl \quad \quad \quad \textbf{endif}\\
	\nl \quad \quad \textbf{endfor}\\
	\nl \textbf{endfor}\\
	\nl \textbf{foreach} $r$ $\code{in}$ $R(\sigma)$ \textbf{do}\\
	\nl \quad \quad \textbf{if} $\neg \exists (d_1,...,d_n) \in M(p) : p \in P_{F} \land \code{color}(p) = \code{color}(r) \land \code{id}(r) = d_1$ \textbf{then}\\
	\nl \quad \quad \quad $\code{registerDeviation}(\code{NONPROPER}\_\code{TERMINATION})$\;
	\nl \quad \quad \quad $\code{jump}(\code{id}(r), p)$\;
	\nl \quad \quad \quad $\code{j} \gets \code{j} + 1$\;
	\nl \quad \quad \textbf{endif}\\
	\nl \textbf{endfor}\\
	\nl $\code{consumeAllObjectsFromSinkPlaces}(P_{F},R(\sigma))$\;
	\nl $\code{k} \gets \code{k} + | R(\sigma) |$\;
	\nl \textbf{return} $(\code{j},\code{k})$\;
	\caption{{\bf Object-Centric Replay with CPNs}}
	\label{alg:alg1}
}
\end{algorithm}

\noindent Let us consider again the replay of $\sigma$ in Table 1 on the CPN of Fig. \ref{fig:cpn-model}. Let us assume that events $e_1$,...,$e_4$ were processed with no deviations detected. Now, consider $e_5 = (\code{new sell order}, \{ (\code{s2}, \code{3}, \code{19.0}, \code{1}) \})$ which implies to fire transition $t_4$ consuming token with id. $\code{s2}$. In the current model marking, however, $\code{s2}$ is not in place $p_4$, but in $p_2$. To execute the model according to $e_5$, token $s_2$ jumps to place $p_4$ as depicted in Fig. \ref{fig:deviation1}. This deviation relates to a sell order that was placed in the order book, but that illegally skipped activity $\code{submit sell order}$.

\begin{figure}[H]
\centering
\includegraphics[scale=0.77]{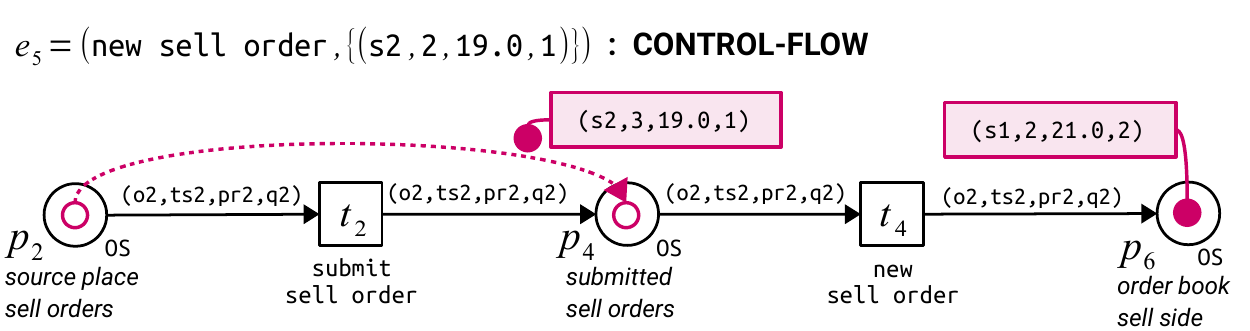}
\caption{Control-flow deviation: $\code{s2}$ is not in $p_4$, so a jump is done to force replay.}
\label{fig:deviation1}
\vspace*{-\baselineskip}
\end{figure}

\begin{figure}[H]
\centering
\includegraphics[scale=0.77]{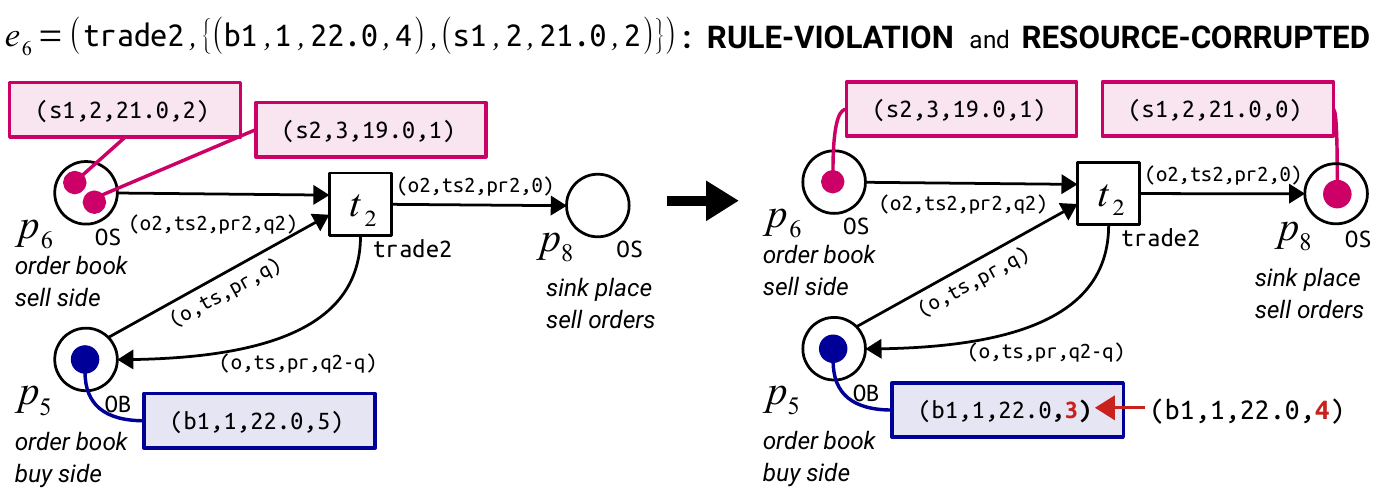}
\caption{Example of a priority rule violation and a resource corruption. }
\label{fig:deviation23}
\vspace*{-\baselineskip}
\end{figure}

\noindent Prior to each transition firing, the method checks if each token to consume is the one that must be selected according to a \emph{priority rule}. To this end, we shall assume that input CPNs may have priority rules on some transitions. Let $b$ a selected binding to fire a transition $t$. We define a priority rule on $t$ as $\Upphi(t) = \bigwedge_{\forall p \in \prescript{\bullet}{}{t}} \upphi_p(M(p),b(\mathcal{E}(p,t)))$, s.t. $b(\mathcal{E}(p,t))$ is the token to consume from input place $p$, and $\upphi_p(M(p),b(\mathcal{E}(p,t)))$ is a priority local rule on $p$; $\upphi_p(M(p),b(\mathcal{E}(p,t)))$ holds if $b(\mathcal{E}(p,t))$ must be consumed before other tokens in $M(p)$. Algorithm $1$ checks the truth value of $\Upphi(t)$ by checking if the local rule of each input place $p$ of $t$ is violated, i.e., in line $11$, function $\code{priorityRuleViolation}((d_1,...,d_n),M(p))$ evaluates to true iff $\Upphi(t)$ is defined and $\upphi_p(M(p),(d_1,...,d_n))$ does not hold. If the function returns true, then a \emph{priority rule violation} is registered as token $(d_1,...,d_n)$ should not have been consumed before other tokens in $p$. For example, let us assign $\Upphi(t) = \upphi_{\code{BUY}}(M(p_5),r_1) \land \upphi_{\code{SELL}}(M(p_6),r_2)$ to transitions $t_5$, $t_6$, and $t_7$ (trade activities) in the CPN of Fig. \ref{fig:cpn-model}, such that:

\medskip

\begin{footnotesize}
\noindent $ \upphi_{\code{BUY}}(M(p_5),r_1) = \forall_{\scaleto{(\code{o},\code{ts},\code{pr},\code{q}) \in M(p_5) \; {\code{id}}(r_1) \neq \code{o}}{8.7pt}} : (\code{price}(r_1) > \code{pr})$

\smallskip

$\quad\quad\quad\quad\quad\quad\quad \lor \; (\code{price}(r_1) = \code{pr} \; \land \; \code{tsub}(r_1) < \code{ts}) $

\smallskip

\noindent $ \upphi_{\code{SELL}}(M(p_6),r_2) = \forall_{\scaleto{(\code{o},\code{ts},\code{pr},\code{q}) \in M(p_6) \; {\code{id}}(r_2) \neq \code{o}}{8.5pt}} : (\code{price}(r_2) < \code{pr})$

\smallskip

$\quad\quad\quad\quad\quad\quad\;\; \lor \; (\code{price}(r_2) = \code{pr} \; \land \; \code{tsub}(r_2) < \code{ts}) $

\end{footnotesize}

\smallskip

\noindent where $r_1$ and $r_2$ are buy and sell orders to consume; the local rule $\upphi_{\code{BUY}}$ on place $p_5$ states that $r_1$ must be the order with the highest price (or with the earliest submitted time if other orders have the same price). The local rule $\upphi_{\code{SELL}}$ on $p_6$ is defined similarly, but $r_2$ must be the order with the lowest price. Let us consider event $e_6$ in Fig. \ref{fig:deviation23}: the rule on $p_6$, to prioritize sell orders with the lowest price, is violated as order $\code{s1}$ with price $\code{21.0}$ is consumed before $\code{s2}$ with price $\code{19.0}$.

\smallskip

\noindent After firing a transition according to an event, we search for \emph{resource corruptions}. Specifically, we check if the values of every transferred token are equal to the values of corresponding objects in the event; this detects if a system transformed object attributes as expected, e.g., in Fig. \ref{fig:deviation23}, after the trade of $\code{1}$ stock between $\code{b1}$ and $\code{s1}$, the stock quantity of $\code{b1}$ decreased from $\code{5}$ to $\code{3}$; however, event $e_6$ shows that the $\code{b1}$'s stocks changed to $\code{4}$, indicating that $\code{b1}$ was corrupted; in case of these deviations, values of the corrupted token are updated according to the values of its corresponding object in the event, e.g., in Fig. \ref{fig:deviation23}, $\code{b1}$'s stocks change to $\code{4}$.

\noindent After replaying a trace, we check \emph{non-proper termination}, that is, whether the system did not fully process all objects. We check if all objects reside in their corresponding sinks. After the replay of the trace in Table 1 on the CPN of Fig. 1, orders $\code{b1}$ and $\code{s2}$ did not arrive at their sinks. These are orders that were not fully handled by the trading system. For these deviations, the method moves these tokens at their sinks, increasing the counter of token jumps $\code{j}$. When all tokens are in the sinks, they are consumed by the “environment”, and the counter of transfers $k$ increases by the number of tokens consumed. Finally, the ratio $1 - \code{j} / \code{k}$ can be used as a fitness metric to measure the extent to which a system (as observed in the trace) complies with the CPN, e.g., if the result of such ratio is $1$, then all behavior observed in the trace complied with the model.

\end{section}

\begin{section}{Prototype and Experiments}

We developed a software prototype\footnote[1]{https://github.com/jcarrasquel/hse-uamc-conformance-checking} of the method proposed using SNAKES \cite{SNAKES2}, a Python library for simulating CPNs. We aimed at detecting deviations within a subset of order books in a real trading system. We considered order books with only \emph{day limit orders}, orders that trade stocks at a fixed price, and that must trade or cancel by the end of a day. The orders considered are not amended after their submission. The system expected behavior is described by the CPN of Fig. \ref{fig:cpn-model}. The method takes as input the CPN of Fig. \ref{fig:cpn-model} and a log where traces relate to the handling of order books during a day. The log was extracted from a set of Financial Information Exchange (FIX) protocol messages \cite{FIX}.\\The messages were exchanged by users and the system during a day, informing activities executed and the status of orders. The set consists of $552935$ FIX messages, whereas the log obtained from such set consists of $73$ traces (order books) and $2259$ events, with a mean of $30.94$ events per trace. A fragment of the deviations file computed by the method is shown in Fig. \ref{fig:deviations}. The file lists deviations detected in events of different traces of the input log. Each line describes precise information of a deviation in the real system: the trace (order book), event number, timestamp, and activity where the error occurred, the object affected, the kind of deviation detected, and an automatically generated description. In this experiment, most of the deviations relate to corruption of orders when executing trades: the prices of some orders changed upon the execution of trades, e.g., in event 1781 the price of the order with id. \code{bSovX} changed from 105 to 100 after trading, and such transformation is not described in the CPN. Thus, this information about deviations can be used by experts to confirm if this is a failure in the system, or instead the model should be slightly refined.

\begin{figure}[H]
\centering
\includegraphics[width=\textwidth]{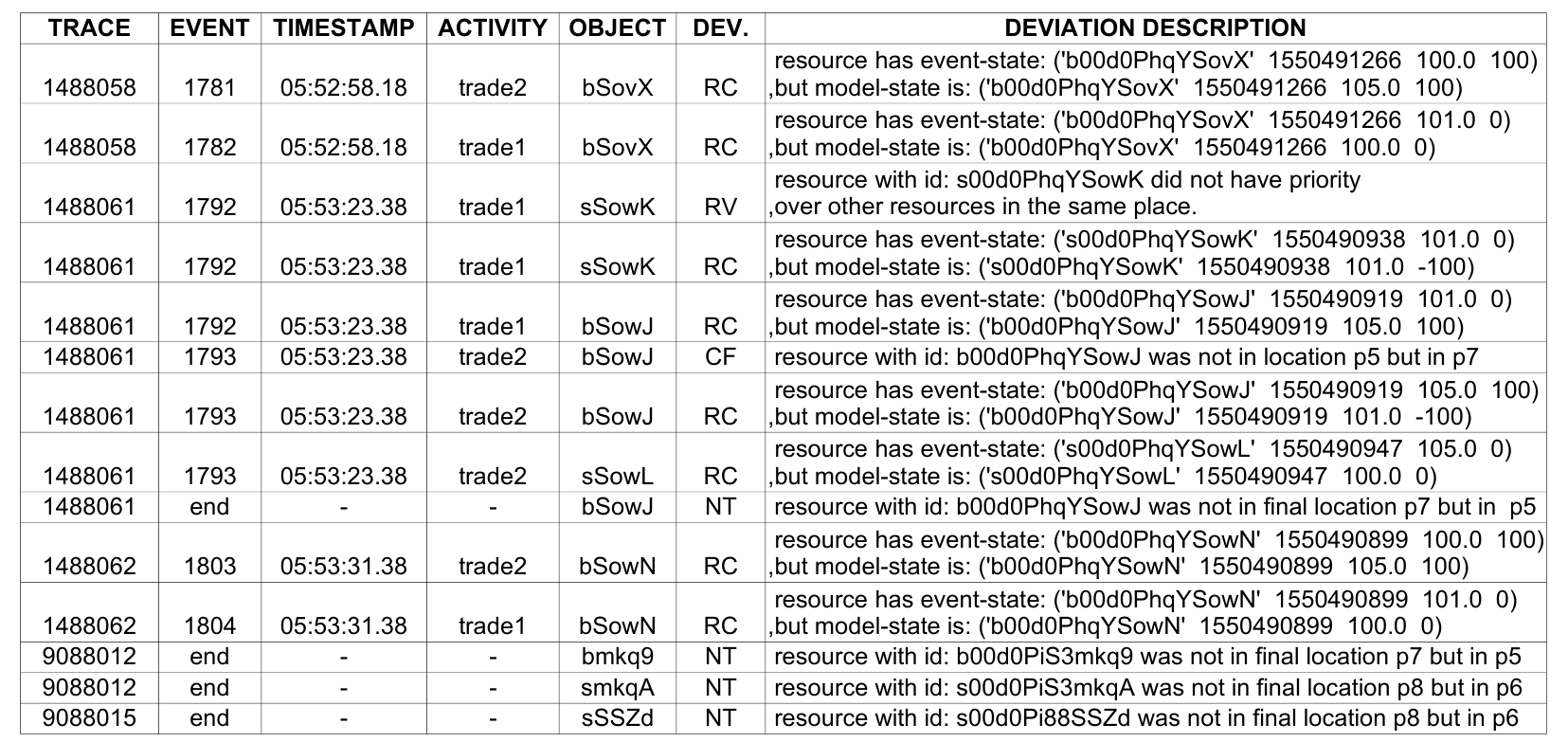}
\vspace*{-\baselineskip}
\caption{Fragment of deviations detected (\code{DEV}): resource corruptions (\code{RC}), priority rule violations (\code{RV}), control-flow deviations (\code{CF}), non-proper termination (\code{NT})}.
\vspace*{-\baselineskip}
\label{fig:deviations}
\end{figure}

\noindent In a second experiment, we show how information obtained during replay, about token jumps and transfers, can be used to enhance an input CPN for visualizing deviations. Using SNAKES, we built a model representing a trading system, similar to the CPN of Fig. \ref{fig:cpn-model}, but with some undesired behavior that shall be uncovered as control-flow deviations: orders may skip activities \code{submit buy order} and \code{submit sell order}, e.g., this may represent malicious users submitting unverified orders via back-doors. Also, activity \code{new sell order} may lead some orders to a deadlock. As input for our method, we consider the model of Fig. \ref{fig:cpn-model} and an artificial event log, that records the system's behavior. The log was generated by our solution, running the CPN that represents the faulty system. The log consists of $100$ traces and $4497$ events, with an average of $44.97$ events per trace. In each trace, there is an average of $10$ buy orders and $10$ sell orders.

\vspace*{-\baselineskip}
\begin{figure}[H]
\centering
\includegraphics[scale=0.74]{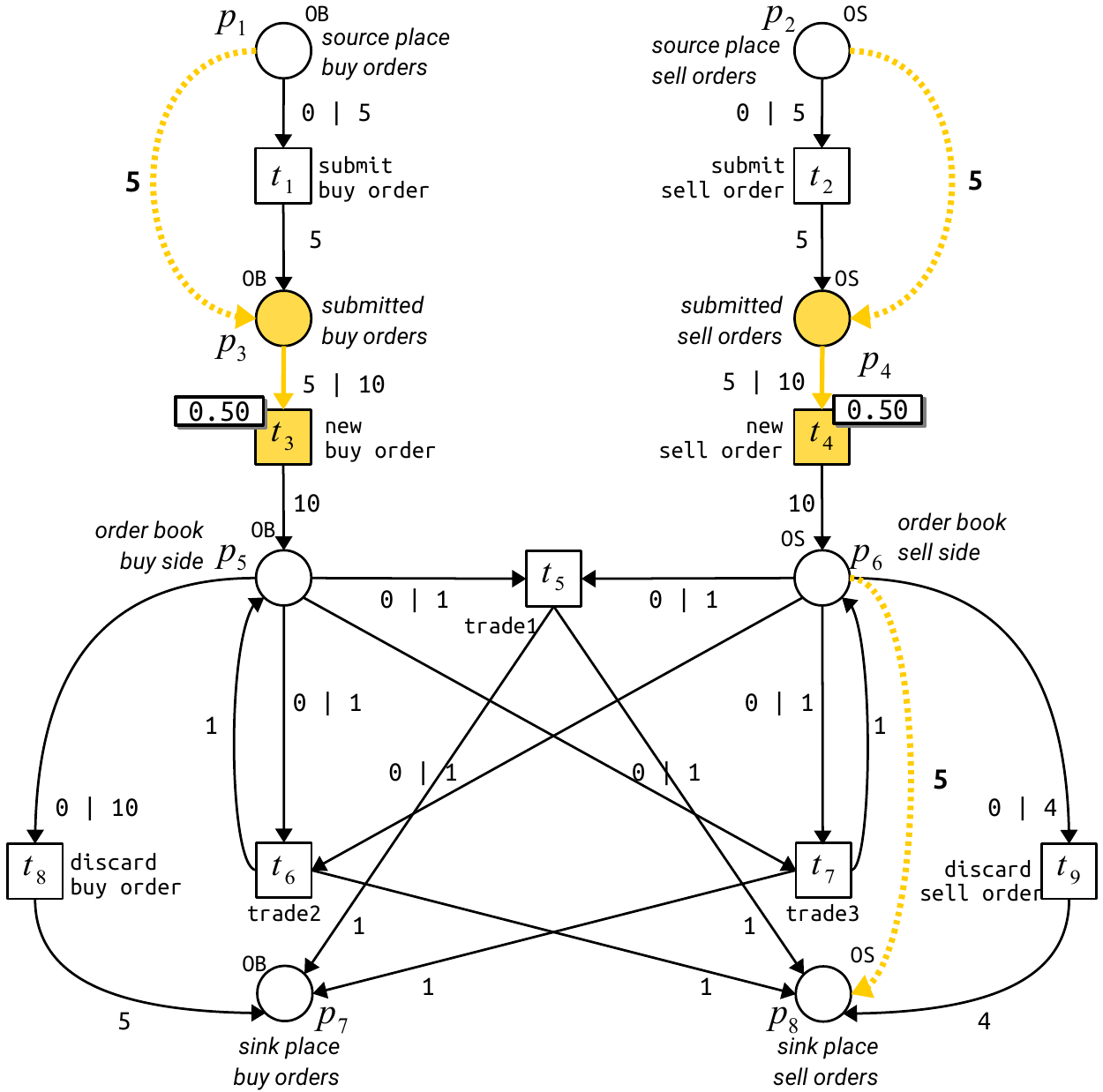}
\caption{Specification model extended with diagnostics computed by our method.}
\vspace*{-\baselineskip}
\label{fig:extendedcpn}
\end{figure}

\noindent Upon the execution of the method, control-flow deviations are detected and reveal the undesired behavior previously described. When detecting such deviations, tokens jump between places via unforeseen model paths in order to continue the replay. Information about token jumps in each place of the CPN, as well as token transfers are registered by our solution. Fig. 5 illustrates how such information is used to enhance the input CPN model. Dotted lines represent token jumps related to the deviations mentioned: jumps from $p_1$ to $p_3$, and from $p_2$ to $p_4$ are from orders that illegally skipped activities \code{submit buy order} and \code{submit sell order}. Also, jumps from $p_6$ to $p_8$ relate to orders that got locked after executing \code{new sell order}. The method detects such locked orders when checking non-proper termination. Input arcs and dotted lines indicate the (rounded) average number of transferred/jumped tokens, considering all log traces. The software prototype tracks the proportion of token transfers/jumps flowing through model components. \emph{Local conformance metrics} are computed using such proportions to measure how deviations affect precise system parts. For example, \code{new buy order} has a measure of $0.5$, meaning that $5$ out of $10$ objects processed by the activity complied with the model path. We refer to \cite{Carrasquel2021} for formal definitions and a further discussion about these local measures.

\end{section}

\begin{section}{Conclusions}

In this paper, we presented a conformance method to search for deviations in trading systems. Different deviations are detected by replaying a system's trace on a CPN. We validated the management of orders in a real system and revealed precise deviations. Another experiment showed how conformance diagnostics can be added to a CPN to display control-flow deviations. A direction for further research may study how to visualize more complex deviation patterns.
\end{section}

\bibliography{references}

\end{document}